\documentclass[conference]{IEEEtran}
\IEEEoverridecommandlockouts
\usepackage{cite}
\usepackage{url}
\makeatletter
\newcommand*{\rom}[1]{\expandafter\@slowromancap\romannumeral #1@}
\makeatother
\usepackage{amsmath,amssymb,amsfonts}
\usepackage{algorithmic}
\usepackage{graphicx}
\usepackage{textcomp}
\usepackage{xcolor}
\def\BibTeX{{\rm B\kern-.05em{\sc i\kern-.025em b}\kern-.08em
    T\kern-.1667em\lower.7ex\hbox{E}\kern-.125emX}}
\begin{document}

\title{Improving EEG based Continuous Speech Recognition\\
{
}
\thanks{}
}

\author{\IEEEauthorblockN{Gautam Krishna}
\IEEEauthorblockA{\textit{Brain Machine Interface Lab} \\
\textit{The University of Texas at Austin}\\
Austin, Texas \\
}
\and
\IEEEauthorblockN{Co Tran}
\IEEEauthorblockA{\textit{Brain Machine Interface Lab} \\
\textit{The University of Texas at Austin}\\
Austin, Texas \\
}
\and
\IEEEauthorblockN{Mason Carnahan}
\IEEEauthorblockA{\textit{Brain Machine Interface Lab} \\
\textit{The University of Texas at Austin}\\
Austin, Texas \\
}
\and
\IEEEauthorblockN{Yan Han}
\IEEEauthorblockA{\textit{Brain Machine Interface Lab} \\
\textit{The University of Texas at Austin}\\
Austin, Texas \\
}
\and
\IEEEauthorblockN{Ahmed H Tewfik}
\IEEEauthorblockA{\textit{Brain Machine Interface Lab} \\
\textit{The University of Texas at Austin}\\
Austin, Texas  \\
}
}

\maketitle

\begin{abstract}
In this paper we introduce various techniques to improve the performance of electroencephalography (EEG) features based  continuous speech recognition (CSR) systems. A connectionist temporal classification (CTC) based automatic speech recognition (ASR) system was implemented for performing recognition. We introduce techniques to initialize the weights of the recurrent layers in the encoder of the CTC model with more meaningful weights rather than with random weights and we make use of an external language model to improve the beam search during decoding time. 

We finally study the problem of predicting articulatory features from EEG features in this paper. 
\end{abstract}


\section{Introduction}
Automatic speech recognition (ASR) system maps acoustic features to text. ASR systems forms front end or back end in many state of the art voice assistant systems like Bixby, Alexa,Siri,Cortana etc. Most of the current state of the art ASR systems are trained only with acoustic features and can operate only with acoustic input, this limits technology accessibility for people who can't speak at all or for people with speaking disabilities like stuttering. On the other hand electroencephalograpgy (EEG) is a non invasive way of measuring electrical activity of human brain. EEG sensors are placed on the scalp of the subjects to obtain the EEG recordings. Recently in \cite{krishna2019speech} authors demonstrated isolated speech recognition using EEG features on a limited English vocabulary of four words and five vowels. In \cite{krishna20} authors demonstrated continuous speech recognition using connectionist temporal classification (CTC) \cite{graves2006connectionist} and attention model \cite{chorowski2015attention} on an English vocabulary of 20 unique sentences using EEG and combination of EEG, acoustic features as input. In \cite{krishna2019state} authors demonstrated continuous speech recognition using different EEG feature sets.  In \cite{krishna20,krishna2019state} authors used state of the art end-to-end ASR models to directly map EEG features to text. 

In \cite{kirchhoff2002combining} authors demonstrated combining articulatory features with acoustic features improves the performance of ASR systems and in \cite{krishna2019speech} authors demonstrated combining EEG features with acoustic features also improve the performance of ASR systems operating in presence of background noise. In \cite{krishna2019state,krishna20} authors didn't take into account the articulatory features for performing speech recognition. In this paper we show that the articulatory features can be used to design more robust ASR EEG encoder models and can help in improving the performance of EEG based continuous speech recognition systems. 

In \cite{krishna2019state,krishna20} authors initialized the weights of the recurrent neural network (RNN) encoder in their CTC network with random weights. In this paper we demonstrate that initializing the weights of the first few RNN layers in the CTC network with weights of a RNN trained to predict concatenation of acoustic and articulatory features from EEG features will help in significantly improving the performance of EEG based speech recognizer.  We further demonstrate predicting articulatory features from EEG features using temporal convolutional network (TCN) \cite{bai2018empirical} model. 

In \cite{krishna20,krishna2019state} authors didn't use external language model during inference time. In this paper we demonstrate that using an external language model during inference time significantly improves the beam search decoder performance of EEG based speech recognizer. Finally in this paper we demonstrate EEG based speech recognition results for a larger English vocabulary size than the ones used by authors in \cite{krishna20,krishna2019state}.

\section{Connectionist Temporal Classification (CTC)}
The encoder of our CTC network consists of two gated recurrent unit (GRU) \cite{chung2014empirical} layers with 128 hidden units and 64 hidden units respectively connected to a temporal convolutional network (TCN) \cite{bai2018empirical} layer with 32 filters as shown in Figure 1. The kernel size for TCN layer was 2, number of stacks of residual blocks was one, padding type was casual and linear activation was used. Batch normalization was not used in the residual blocks when the model was used with data set A but it was applied when used with data set B. Also a dropout regularization with a dropout rate 0.1 was applied in the TCN layer when the model was used with data set B. 
The GRU layers contained dropout regularization with a dropout rate of 0.1. 
The decoder of the CTC network consists of a combination of dense layer and softmax activation. The output of the encoder layer is fed into the decoder at every time step. The two GRU layers in the encoder network are initialized with weights of the GRU layers of the model shown in Figure 2. The model in Figure 2 consists of two layers of GRU with 128 and 64 hidden units respectively connected to a time distributed dense layer of 19 hidden units with linear activation to predict concatenation of acoustic or Mel-frequency cepstrum coefficients (MFCC) features of dimension 13 and articulatory features of dimension 6 or a net feature dimension of 19 at every time step. The model shown in Figure 2 was trained for 500 epochs with mean squared error (MSE) as loss function with adam optimizer \cite{williams1989learning} and batch size one. Basically the model shown in Figure 2 is a GRU based regression model which predicts combination of acoustic and articulatory features from EEG features of dimension 30 at every time step. 

The CTC model was trained for 120 epochs to optimize the CTC loss using adam optimizer. More details of CTC loss function are described in \cite{graves2006connectionist,graves2014towards,krishna20,krishna2019state}. The batch size was set to 32 for the CTC model. The TCN layer in the CTC encoder was initialized with random weights. The main motivation behind this idea was that the first two GRU layers of the CTC encoder will help in discovering acoustic and articulatory representations from the input EEG features and the TCN layer will learn the mapping of those representations to text. There was no fixed value for the time steps for the encoder of the CTC model. As usual the number of time steps is equal to the product of the sampling frequency
of the input features and input sequence length. During inference time of the CTC model we used a combination of CTC beam search decoder and an external 4-gram language model. Along with beam search we include the log of the probability assigned by the language model for the label sequence during inference time. This technique is commonly known in ASR literature as shallow fusion \cite{toshniwal2018comparison}. 
We used a character based CTC model for this work. The CTC model predicted a character at every time step. The model (Figure 1) takes \textbf{only EEG as input during training and test time}. 
Figure 3 shows the loss convergence of the CTC ASR model and Figure 4 shows the loss convergence of the GRU regression model described in Figure 2. All the scripts were written using python keras deep learning and tensorflow 2.0 framework. 

\begin{figure}[h]
\begin{center}
\includegraphics[height=8.5cm, width=\linewidth,trim={0.1cm 0.1cm 0.1cm 0.1cm}]{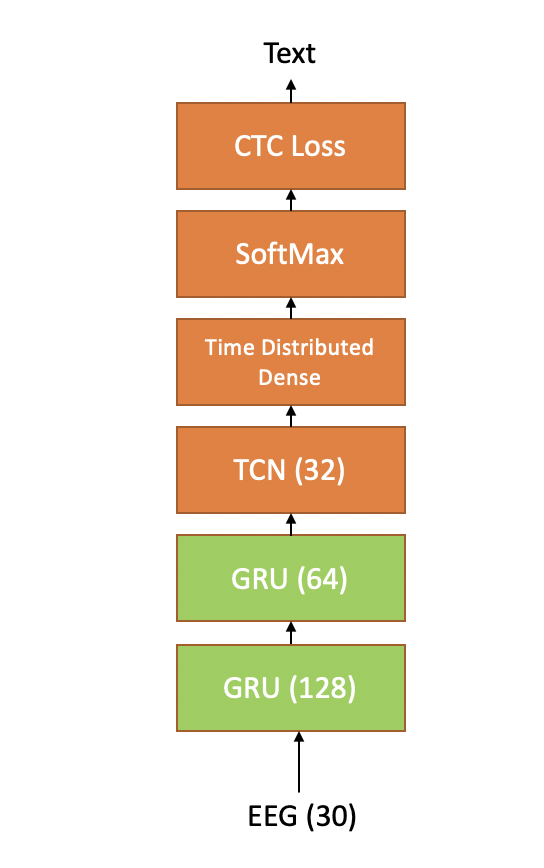}
\caption{CTC ASR Model} 
\label{1vsall}
\end{center}
\end{figure}

\begin{figure}[h]
\begin{center}
\includegraphics[height=7cm, width=\linewidth,trim={0.1cm 0.1cm 0.1cm 0.1cm}]{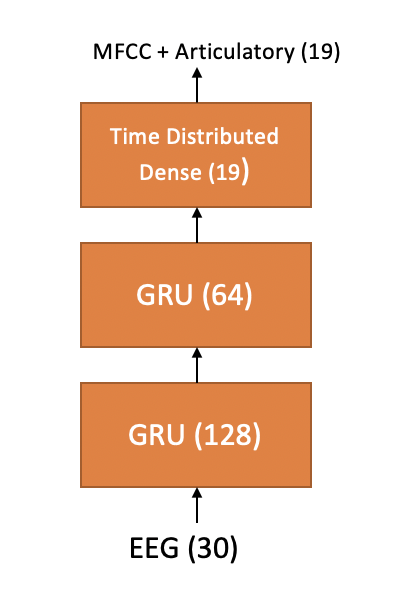}
\caption{Model to derive the initializing weights for the GRU layers of the CTC network} 
\label{1vsall}
\end{center}
\end{figure}

\begin{figure}[h]
\begin{center}
\includegraphics[height=5cm, width=0.4
\textwidth,trim={0.1cm 0.1cm 0.1cm 0.1cm},clip]{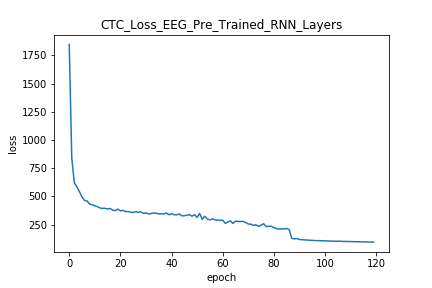}
\caption{CTC loss convergence} 
\label{1vsall}
\end{center}
\end{figure}

\begin{figure}[h]
\begin{center}
\includegraphics[height=5cm, width=0.4
\textwidth,trim={0.1cm 0.1cm 0.1cm 0.1cm},clip]{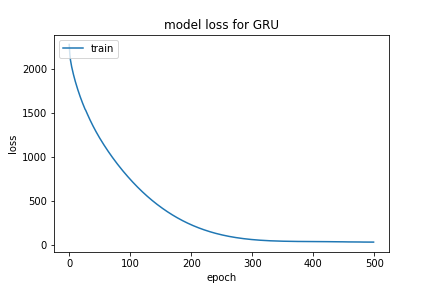}
\caption{GRU regression loss convergence} 
\label{1vsall}
\end{center}
\end{figure}


\begin{figure}[h]
\begin{center}
\includegraphics[height=5cm, width=0.4
\textwidth,trim={0.1cm 0.1cm 0.1cm 0.1cm},clip]{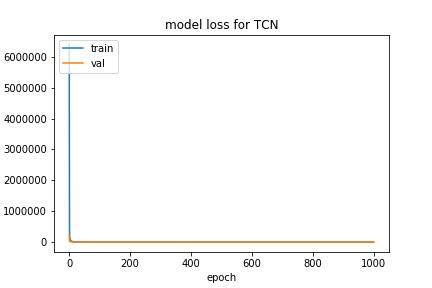}
\caption{TCN regression loss convergence} 
\label{1vsall}
\end{center}
\end{figure}

\begin{figure}[h]
\begin{center}
\includegraphics[height=3cm,width=0.25\textwidth,trim={1cm 1cm 1cm 0.1cm},clip]{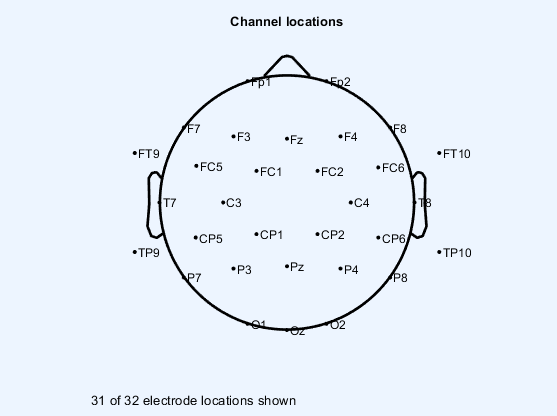}
\caption{EEG channel locations for the cap used in our experiments} 
\label{1vsall}
\end{center}
\end{figure}

\section{Model to predict articulatory features from EEG features} 
For predicting articulatory features of dimension 6 from EEG features of dimension 30 we used a model consisting of a temporal convolutional network (TCN) \cite{bai2018empirical} layer with 128 filters connected to a time distributed dense layer of 6 hidden units with linear activation to predict articulatory features of dimension 6 at every time step. A dropout regularization with dropout rate 0.2 was applied after the TCN layer. The kernel size for TCN layer was 2, number of stacks of residual blocks was one, padding type was casual and linear activation was used. Batch normalization was not used in the residual blocks. 
Mean squared error (MSE) was used as the loss function for this regression model and model was trained for 1000 epochs with adam optimizer. The batch size was set to one.  Figure 5 shows the loss convergence for the model when used with Data set A.

\section{Design of Experiments for building the database}

We used two data sets for this work. The first data set or data set A consists of seven male subjects who took part in our EEG-Speech experiment. All the seven subjects were UT Austin graduate students in their mid twenties. English was not their mother tongue. 

They  were asked to speak the first 30 English sentences from USC-TIMIT database \cite{narayanan2014real} and their simultaneous speech and EEG signals were recorded. This data was recorded in presence of background noise of 65dB. The music played from our lab computer was used as the source of background noise. 
Each subject was asked to repeat the experiment two more times.
The second data set or data set B was the data set A used by authors in \cite{krishna20} where data was recorded in absence of external background noise but a 40 dB noise due to room ventilation fan was observed. Data set B can be approximately considered as a clean data set. 
In both the data sets, the subjects read out loud the English sentences that were shown to them on a computer screen. We used 80 \% of the total data as training set and remaining as test set for all the models for both the data sets. Whereas in \cite{krishna20} authors used data from first 8 subjects from data set A for training the model. The way we splitted data in this work is different from the methods used by authors in \cite{krishna20}. 

We used Brain product's ActiChamp EEG amplifier. Our EEG cap had 32 wet EEG electrodes including one electrode as ground as shown in Figure 6. We used EEGLab \cite{delorme2004eeglab} to obtain the EEG sensor location mapping. It is based on standard 10-20 EEG sensor placement method for 32 electrodes.

\begin{table*}[!ht]
\centering
\begin{tabular}{|l|l|l|l|l|l|l|}
\hline
\textbf{\begin{tabular}[c]{@{}l@{}}Total \\ Number\\ of \\ Sentences\end{tabular}} & \textbf{\begin{tabular}[c]{@{}l@{}}Number of \\  Unique\\ sentences\\ contained\end{tabular}} & \textbf{\begin{tabular}[c]{@{}l@{}}Total Number\\ of words \\ contained\end{tabular}} & \multicolumn{1}{c|}{\textbf{\begin{tabular}[c]{@{}c@{}}Number\\ of \\ Unique words\\ contained\end{tabular}}} & \textbf{\begin{tabular}[c]{@{}l@{}}Number of \\ letters\\ contained\end{tabular}} & \multicolumn{1}{c|}{\textbf{\begin{tabular}[c]{@{}c@{}}WER\\ (\%)\\ GRU \\ layers\\ random\\ weights\\ +\\ LM\end{tabular}}} & \multicolumn{1}{c|}{\textbf{\begin{tabular}[c]{@{}c@{}}WER\\ (\%)\\ GRU\\ layers\\ pre \\ trained\\ weights\\ +\\ LM\end{tabular}}} \\ \hline
21                                                                                 & 5                                                                                             & 134                                                                                   & 29                                                                                                            & 575                                                                               & 82.93                                                                                                                        & \textbf{72.57}                                                                                                                      \\ \hline
42                                                                                 & 10                                                                                            & 277                                                                                   & 59                                                                                                            & 1121                                                                              & 77.66                                                                                                                        & \textbf{75.5}                                                                                                                       \\ \hline
63                                                                                 & 15                                                                                            & 408                                                                                   & 84                                                                                                            & 1891                                                                              & 85.78                                                                                                                        & \textbf{82.5}                                                                                                                       \\ \hline
84                                                                                 & 20                                                                                            & 536                                                                                   & 106                                                                                                           & 2334                                                                              & 86.3                                                                                                                         & \textbf{80.64}                                                                                                                      \\ \hline
105                                                                                & 25                                                                                            & 652                                                                                   & 132                                                                                                           & 2863                                                                              & 97.05                                                                                                                        & \textbf{77.54}                                                                                                                      \\ \hline
126                                                                                & 30                                                                                            & 743                                                                                   & 153                                                                                                           & 3614                                                                              & 103                                                                                                                          & \textbf{87.7}                                                                                                                       \\ \hline
\end{tabular}
\caption{WER on test set for CTC model with GRU layers with random weights vs GRU layers with weights derived from the pre trained EEG to MFCC+Articulatory GRU regression model for Data set A}
\end{table*}

\section{EEG and Speech feature extraction details}

We followed the same EEG and speech preprocessing methods used by authors in \cite{krishna2019speech,krishna20}. 
EEG signals were sampled at 1000Hz and a fourth order IIR band pass filter with cut off frequencies 0.1Hz and 70Hz was applied. A notch filter with cut off frequency 60 Hz was used to remove the power line noise.
EEGlab's \cite{delorme2004eeglab} Independent component analysis (ICA) toolbox was used to remove other biological signal artifacts like electrocardiography (ECG), electromyography (EMG), electrooculography (EOG) etc from the EEG signals. 
We extracted five statistical features for EEG, namely root mean square, zero crossing rate,moving window average,kurtosis and power spectral entropy \cite{krishna2019speech,krishna20}. So in total we extracted 31(channels) X 5 or 155 features for EEG signals.The EEG features were extracted at a sampling frequency of 100Hz for each EEG channel.

The recorded speech signal was sampled at 16KHz frequency. We extracted Mel-frequency cepstrum coefficients (MFCC) as features for speech signal.
We extracted MFCC features of dimension 13. The MFCC features were also sampled at 100Hz, same as the sampling frequency of EEG features. For training the model explained in Figure 2, for Data Set B we directly used the MFCC features extracted from Data Set B recorded clean speech and for Data set A we observed similar performance when we directly used MFCC features extracted from Data Set A recorded noisy speech and after speech enhancement as the articulatory features used with MFCC were noise robust. 

We used acoustic-to-articulatory speech inversion tool introduced by authors in \cite{sivaraman2016vocal,seneviratne2018noise} to extract articulatory features of dimension 6 from the recorded speech signal. The articulatory features were also extracted at the same sampling frequency of 100 Hz as that of the MFCC and EEG features. The six articulatory tract variables (TV's) that were extracted were Lip Aperture (LA), Lip Protrusion (LP), Tongue Body Constriction Location (TBCL), Tongue Body Constriction Degree (TBCD), Tongue Tip Constriction Location (TTCL) and Tongue Tip Constriction Degree (TTCD) \cite{sivaraman2016vocal,seneviratne2018noise}. The articulatory features estimated were robust to background noise \cite{seneviratne2018noise}. 
\section{EEG Feature Dimension Reduction Algorithm Details}
After extracting EEG and acoustic features as explained in the previous section, we used Kernel Principle Component Analysis (KPCA) \cite{mika1999kernel} to denoise the EEG feature space as explained by authors in \cite{krishna20,krishna2019speech}. 
We reduced the 155 EEG features to a dimension of 30 by applying KPCA for both the data sets. We plotted cumulative explained variance versus number of components to identify the right feature dimension. We used KPCA with polynomial kernel of degree 3 \cite{krishna2019speech,krishna20}. 

\section{Results}
We used word error rate (WER) as performance metric to evaluate the performance of the CTC ASR model on test set data for various number of sentences as shown in Table 1 and 2. The average WER is reported in Table 1,2. Language model (LM) was included during inference time. Without the language model we observed higher error rates than the ones reported in Table 1,2. We observed roughly around 5 to 10 \% increase in error rates for most of the experiments when the external language model was not included during inference time. Table 1 shows the test time result for data set A and table 2 shows the test time result for data set B. 
As seen from Table 1,2 results it is quite evident that initializing the GRU layers in the encoder of the CTC network with pre trained regression model weights significantly improves the test time performance of EEG based CTC speech recognizer especially as the vocabulary size increase. 

We used two performance metric to evaluate the performance of TCN regression model for predicting articulatory features from EEG features. The two performance metrics were root mean squared error (RMSE) and normalized RMSE between the predicted articulatory features during test time and ground truth articulatory features from test set. The RMSE values were normalized by dividing the RMSE values with the absolute difference between the maximum and minimum value in the test set observation vector. We observed an average RMSE of \textbf{0.632} and average normalized RMSE of \textbf{0.115} on the test set with data set A and an average RMSE of \textbf{1.30} and average normalized RMSE of \textbf{0.238} on the test set with data set B . 
\begin{table}[!ht]
\centering
\begin{tabular}{|l|l|l|l|}
\hline
\textbf{\begin{tabular}[c]{@{}l@{}}Total \\ Number\\ of \\ Sentences\end{tabular}} & \textbf{\begin{tabular}[c]{@{}l@{}}Total\\  Number\\ of \\ words \\ contained\end{tabular}} & \multicolumn{1}{c|}{\textbf{\begin{tabular}[c]{@{}c@{}}WER\\ (\%)\\ GRU \\ layers\\ random\\ weights\\ +\\ LM\end{tabular}}} & \multicolumn{1}{c|}{\textbf{\begin{tabular}[c]{@{}c@{}}WER\\ (\%)\\ GRU\\ layers\\ pre \\ trained\\ weights\\ +\\ LM\end{tabular}}} \\ \hline
30                                                                                 & 200                                                                                         & 82.63                                                                                                                        & \textbf{74.36}                                                                                                                      \\ \hline
60                                                                                 & 403                                                                                         & 84.30                                                                                                                        & \textbf{74.45}                                                                                                                      \\ \hline
90                                                                                 & 600                                                                                         & 82.67                                                                                                                        & \textbf{77.76}                                                                                                                      \\ \hline
120                                                                                & 773                                                                                         & 88.94                                                                                                                        & \textbf{79.68}                                                                                                                      \\ \hline
150                                                                                & 948                                                                                         & 90.39                                                                                                                        & \textbf{81.97}                                                                                                                      \\ \hline
180                                                                                & 1112                                                                                        & 85.39                                                                                                                        & \textbf{84.9}                                                                                                                       \\ \hline
\end{tabular}
\caption{WER on test set for CTC model with GRU layers with random weights vs GRU layers with weights derived from the pre trained EEG to MFCC+Articulatory GRU regression model for Data set B. Number of unique sentences and number of unique words for each row is same as that of Table 1}
\end{table}

We further observed that for data set A, results for larger vocabulary size can be slightly improved by adding a couple of GRU layers taken from a pre-trained MFCC + articulatory to text ASR model, to the CTC encoder model described before in Figure 1. The new GRU layers were added after the GRU(64) layer and the new GRU layers were frozen during training of the model. The output of the new pair of GRU layers are passed to the TCN layer. The TCN layer as usual was initialized with random weights.

The intuition behind adding these GRU layers was since this layers were borrowed from a CTC ASR model which was trained to predict text from MFCC + articulatory features, these GRU layers might help in transforming the MFCC and articulatory representations learned by GRU (64) from EEG into features which which can be easily translated to text by TCN(32) layer. By adding these additional layers we observed a lower WER of \textbf{78.39 \%, 78.9 \% and 85.33 \%} for number of unique test sentences 15, 20 and 30 respectively.  The new pair of GRU layers contained 128 and 64 hidden units respectively. Since we observed only a slight improvement in results for larger vocabulary for data set A, we didn't perform experiments using these new additional set of GRU layers for data set B. 

In \cite{krishna2019speech} authors demonstrated that EEG sensors T7 and T8 contributed most towards test accuracy for EEG based isolated speech recognition.  Similarly in \cite{flinker2015redefining,chartier2018encoding} authors demonstrated that ventral sensorimotor cortex (vSMC), superior temporal gyrus (STG) and inferior frontal gyrus (IFG) plays crucial role in human speech production and perception. This regions belongs to the frontal and temporal lobes in human brain. Because of the source localization issue associated with EEG recordings, it is impossible to find the exact EEG sensor placements to get the  electrical activities originating from STG or IFG or vSMC, hence we tried performing ASR experiments using the same ASR model described in Figure 1 using data set B on a test set vocabulary of 180 sentences consisting of 30 unique sentences first with the features (dimension 20, five features per channel, no dimension reduction performed here) from all the temporal lobe EEG sensors ( T7,T8,TP9,TP10) and observed a WER of \textbf{86.52 \%} and then with features(dimension 65) from all frontal lobe EEG sensors ( F3,F4,F7,F8,FC1,FC2,FC5,Fp1,Fp2,FT9,FT10,Fz) and observed a WER of \textbf{85.45 \%} and finally by combining features from all the temporal and frontal lobe EEG sensors of dimension 85 and observed a WER of \textbf{85.32 \%} but all these WER's were higher than \textbf{84.9 \%} reported in Table 2 where features from all EEG sensors and dimension reduction was used. 
\section{Conclusion and Future work}
In this paper we demonstrated various techniques to improve the performance of EEG features based continuous speech recognition systems. We further demonstrated predicting articulatory features from EEG features with very low RMSE and normalized RMSE. 

For future work we plan to conduct experiments with data collected from subjects with speaking disabilities and build a larger speech-EEG data base.

\section{Acknowledgement} 
We would like to thank Kerry Loader and Rezwanul Kabir from Dell, Austin, TX for donating us the GPU to train the models used in this work. The first author would like to thank Prof Alex Dimakis from ECE department at UT Austin for suggesting us to carry out experiments using TCN layers.

\bibliographystyle{IEEEtran}

\bibliography{refs}
\end{document}